\documentclass[twocolumn,preprint,3p]{elsarticle}

\usepackage{graphicx}
\usepackage{graphicx,epstopdf}
\usepackage{xcolor}
\usepackage{dcolumn}
\usepackage{bm}
\usepackage{mathrsfs} 
\usepackage{hyperref}
\usepackage{amsmath}
\usepackage{amssymb}
\usepackage[utf8]{inputenc}
\usepackage[T1]{fontenc}
\usepackage{mathptmx}
\usepackage{mathtools}
\usepackage{etoolbox}
\usepackage[export]{adjustbox}
\usepackage{euscript}
\usepackage{tikz}
\usepackage{relsize}
\usepackage{soul}
\journal{Chaos, Solitons \& Fractals}

\begin{document}

\title{Optimizing the location of the colony of foragers with Collective Learning}

\author{Sanchayan Bhowal}
\address{Indian Statistical Institute, Bangalore Centre, 8th Mile, Mysore Road, Bangalore- 560 059 India}
\ead{sanchayan.bhowal2509@gmail.com}

\author{Ramkrishna Jyoti Samanta}
\address{Indian Statistical Institute, Bangalore Centre, 8th Mile, Mysore Road, Bangalore- 560 059 India}
\ead{akashnilsamanta@gmail.com}

\author{Arnob Ray}
\address{Physics and Applied Mathematics Unit, Indian Statistical Institute, 203 B. T. Road, Kolkata 700108, India}
\ead{arnobray93@gmail.com}

\author{Sirshendu Bhattacharyya}
\address{Department of Physics, Raja Rammohun Roy Mahavidyalaya, Radhanagar, Hooghly 712406, India}

\author{Chittaranjan Hens}
\address{Center for Computational Natural Science and Bioinformatics, International Institute of Informational Technology, Gachibowli, Hyderabad-500032, India}

\date{\today}

\begin{abstract}
Animal groups collaborate with one another throughout their lives to better comprehend their surroundings. Here, we try to model, using continuous random walks, how the entire process of birth, reproduction, and death might impact the searching process. We attempt to simulate an ecosystem where the post-reproductive foragers leave their colonies to discover where the targets are while others stay and breed at the base. 
Actually, a group of foragers searches for a location from where they access the targets for food supply. Particularly, we have explored a hypothetical situation in which the relocation to the new position depends on the agreement level of the species as well as an additional waiting time due to this agreement level. In this backdrop, detailed numerical results reveal that searching for an optimal position at an optimal mean time can be captured for a suitable range of the agreement level. We have also shown, for a given agreement level,  the optimal mean time linearly increases with the Death-to-Birth ratio. 
\end{abstract}



\maketitle


\section{\label{sec:level1}Introduction}
Many animals rely extensively on learning mechanisms to adapt to their environment. The movement of animal groupings is primarily motivated by a set of objectives (mostly food). Many animal species live in groups and work together to attain goals\ \cite{success, Collective_foraging}. On the other hand, these objectives may be scattered and too far away from the initial location of the colony. In this situation, the groups frequently strive to relocate their colonies to locations that are close to all of the objectives\ \cite{relocation}. Many well-known benefits of socialization include reduced predation risk and increased sensing and decision-making abilities while foraging for food in unfamiliar surroundings\ \cite{predation}. 

\par For instance, female whales in their reproductive years lead groups during collective migration in salmon feeding sites\ \cite{whale-1}. Leadership by the post-reproductively old is especially noticeable in tough years with low salmon abundance. This discovery is significant because salmon abundance influences both mortality and reproductive success in resident killer whales\ \cite{whale-2,whale-3}. Nest location selection in \textit{Leptothorax albipennis} has also been studied previously\ \cite{ant}. Colonies were discovered to make complex decisions, considering factors other than the fundamental advantages of each location. Leadership qualities of individuals have a considerable impact on the future of a population who help in making decisions whether or not to move based on a specific kind of information they send to the colony\ \cite{lead-1,lead-2}. Several scientists have recently suggested that advanced social insect colonies are higher-order cognitive entities or supraorganismal systems capable of analyzing conditions and developing adaptive solutions to challenges. Honeybee foragers collaborate by exchanging information about plentiful food sources\ \cite{honeybee,honeybee-2}.
\par
Collective learning is suggested to be beneficial for animal groups. They may make a number of decisions based on prior accomplishments during foraging, minimizing the amount of time spent looking for things\ \cite{opt-1,site,opt-2,opt-3}. Markovian random walk models, which assume foragers have no memory, have proved useful in determining how specific kinesthetic awareness and resource distribution effect foraging success\ \cite{memory,signal,fish}. Such models, on the other hand, overlook the fact that animals repeat specific behaviors and are unable to account for the impact of previous movement decisions, so it is often of interest to study the effects of the foragers' interaction network on collective learning\ \cite{collearn,bio}.The information transfer among the group allows them to learn about the global environment. Communication across long range can be observed in many animals such as elephants\ \cite{ele,gira,spi}. We wish to quantify here, by means of uniform Markovian random walks, how the entire process of birth, reproduction, death, and information transfer, which may be instantaneous, can affect the searching process. The structure of social networks is likely to be relevant in such processes since certain individuals are more important than others(the post-reproductive organisms) for transmitting information on food locations. 

\par
Here, we try to hypothetically recreate an ecological setup based on a continuous random walk model. Here, some foragers leave their colonies to learn where targets are, while others stay and breed at the base. The colony is then relocated to the most advantageous positions, where all objectives are within easy reach. Once the population has reached an ideal position, it can stay there for the rest of its existence until a new target emerges. In order to do so, the elders of the community who lost their potential to reproduce leave the colony in search of targets. Once a target is reached, it sends a signal to the colony. Now, the colony must decide whether or not to relocate to the best site. Some foragers agree to relocate, but this may not be enough to persuade the rest of the group to relocate. As a result, they try to convince others, which costs time, and this time acts as a penalty for reaching optimum. The colony shifts if a certain number of individuals agree. 
Our target is to track a suitable set of parameters (agreement level, penalty time, and Death-to-Birth ratio) in order to find out the optimal time to reach the final destination. We have shown that our stochastic simulation, under a proper choice of agreement level, eventually attains the centroid. The paper is organized as follows. In the Sec.\ \ref{sec:model} we have discussed the models, i.e., random movement of the agents in a finite domain. Related parameters are also discussed. All the numerical results are discussed in the Sec.\ \ref{sec:results}. We conclude our results from our intuitive model in Sec.\ \ref{sec:disc}.

\begin{figure*}[!t]
	\centerline{\includegraphics[width=0.9\textwidth]{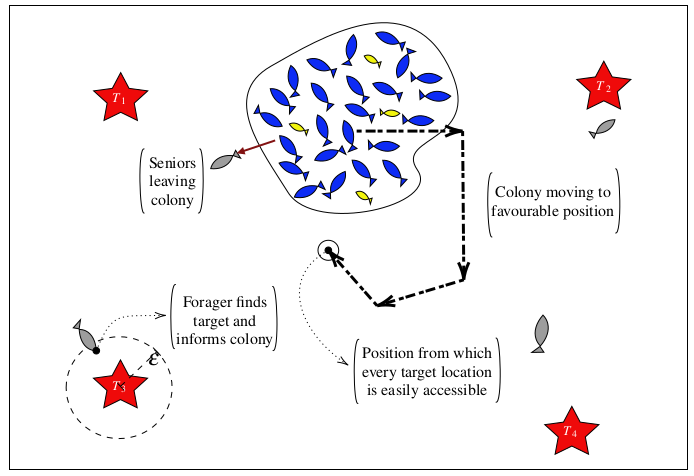}}	
	\caption{{\bf Schematic diagram of the process.} This plot captures the whole process of our model pictorially. The yellow organisms represent infants who are unable to reproduce. The adult individuals who can reproduce are represented by blue organisms, whereas the seniors who have lost their ability to reproduce are represented by grey. As we observe, new infants are being born, and one of the adult organisms ages and loses the ability to reproduce. The aged individual then departs the colony in search of targets. $T_1$, $T_2$, $T_3$, and $T_4$ are four targets which are indicated by red star. A forager detects a target when this forager enters the $\epsilon$ neighborhood of any target. Then, it gives information to the colony. As desired, the colony subsequently updates its location to reach the optimal position marked as a black dot.}
\end{figure*}\label{f1}

\section{ENVIRONMENT AND MODEL DESCRIPTION}
\label{sec:model}

Our model starts with a community of $N$ members ranging in age randomly chosen from one to hundred, setting out on their journey of in search of $N_T$ target points in a two-dimensional space limited by $[-a,a]\times[-a,a]$. The model considers birth, ageing, death and age-related fertility to be the controlling factors in the dynamics. However the gender of the individuals has not been taken into account. Here one simulation step has been taken to be the time unit of the age of the members. In addition to the three parameters such as agreement level $(p)$, birth rate $(B)$, and death rate $(D)$, we introduce another quantity called penalty factor $(\lambda)$ which will be discussed in the description later. The process begins by generating a square of length $\delta$ which is called \emph{Colony}. The centre of the colony is uniformly chosen at random in the $2$D space constrained within $[-a/2,a/2]\times[-a/2,a/2]$. We draw a schematic diagram in Fig.\ \ref{f1}. Here, we observe a colony consists of infant and adult organisms, four targets. This figure inform us how colony of organisms shifts based on their internal interactions.

\par The initial population of $100$ foragers is then randomly positioned inside the colony. Their ages vary from $1$ to $100$, i.e., one such forager per age. Because our model incorporates the processes of birth, aging, and death, foragers above the age of $100$ may emerge. As time passes, the processes give rise to various age groups. Now we try to provide a specific role for each of the three age groups, namely those under the age of $20$, those between the ages of $20$ and $60$, and those exceeding $60$. The foragers under the age of $20$ are neither involved in reproduction nor foraging, foragers with age in between $20$ and $60$ are involved in reproduction and those with age greater than $60$ achieve menopause (loses the ability to reproduce) leave the colony and set out in search of targets (for the whales, it is the salmon-rich areas), as shown in Fig.\ \ref{f1}. As the foragers from age $1$ to $60$ remain in the colony and don't participate in foraging, we call them 'static Foragers'.
In this model, we have tried to fix four target points: $T_1,T_2,T_3$ and $T_4$.

\par All the foragers who leave the colony in search of food follow a rule of movement. While moving, step size and direction play an important role. While the step size is fixed at $h=0.1$, where h is the step size, the direction is chosen based on $\theta$ where $\theta$ is the angle of movement with respect to the positive $x$-axis. Here, $\theta$ is uniformly chosen from $(-\pi,\pi)$. Therefore, $\theta \sim$ Uniform~$(-\pi,\pi)$ for each step. Let the position of the forager be denoted by $(X_n,Y_n)$. Then, the motion of forager is described by \cite{volpe, sar}
\begin{align}
X_{n+1} &=X_{n} + h v_0\cos(\theta_n)\\
Y_{n+1} &=Y_{n} + h v_0\sin(\theta_n)
\end{align}
The explicit distribution for $\cos(\theta_n)$ and $\sin(\theta_n)$ can be determined. Let, $\Delta X_n=X_{n+1}-X_n$ and $\Delta Y_n=Y_{n+1}-Y_n$,
\begin{equation}
\begin{aligned}
\begin{bmatrix}\Delta X_n \\ \Delta Y_n\end{bmatrix} &= \begin{bmatrix}hv_0\cos\theta_n \\ hv_0\sin \theta_n\end{bmatrix}.
\end{aligned}
\end{equation}
Because $\theta_n \sim \text{Uniform}(-\pi,\pi)$, so the distribution function of $\Delta X_n$ and $\Delta Y_n$ are,
\begin{equation}
\begin{aligned}
F_{\Delta X_n}(x) & = \frac{1}{2}+\frac{\arcsin{\frac{x}{hv_0}}}{\pi} \\
F_{\Delta Y_n}(y) & = \frac{1}{2}+\frac{\arcsin{\frac{y}{hv_0}}}{\pi}.
\end{aligned} 
\end{equation}
This shows that $\Delta X_n\sim \Delta Y_n$, i.e. $\Delta X_n$ and $\Delta Y_n$ are identically distributed. Moreover, this brings out that the walk taken by the species is isotropic in nature.

\begin{figure}[h]

	\centerline{\includegraphics[scale=0.65]{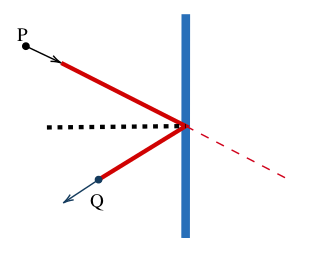}}
	\caption{{\bf Reflective boundary condition.} When the forager encounters the boundary in its path, it is reflected inside the bounded region. At the time of reflection, the forager moves the remaining distance along the direction of reflection.}
	\label{f2}
	
\end{figure}

\par
As already discussed, the space of movement is restricted. As the foragers move, they may hit the boundary multiple times.
As seen in Fig.\ \ref{f2}, whenever a forager comes in contact with a boundary in its path, it is reflected inside the bounded region. At the time of reflection, the orientation of the particle also changes along the direction of reflection. Such conditions are often suitable due to the availability of finite space. This, we call as the \emph{Reflective Condition} in our model\ \cite{refl-1,refl-2,refl-3}. This is all about the movement of foragers.

\par Besides the process of foraging, the birth and death of foragers also play an important role in the process. The static-foragers with ages between $20$ and $60$ are taken to be capable of reproducing. Let the number of new foragers produced at $n$-th step be $\mathcal{X}_n$. Then we assume that $\mathcal{X}_n\sim$ Poisson($\mu$) where $\mu= mB$, where $m$ is the number of foragers who are capable of reproducing at that time step\ \footnote[1]{Note: The position of new foragers are generated randomly inside the colony.}. The foragers' birth process has been purposefully intended to be Poisson$(mB)$. Now, we are going to reason behind the selection of Poisson distribution. Let us assume that the number of mating processes in the colony be $ml$, and $l$ is constant. Each of these processes has the potential to either produce a new forager or fail. Let $r$ be the success probability. As a result, the number of people created follows the Binomial$(ml,r)$ distribution. Now, this is approximated with the Poisson$(mB)$ distribution (as the number of mating processes will be quite large compared to successes), where $B=lr$ is the birth rate assumed to be constant for a population. 
\par
Another factor Death rate ($D$) determines the probability of death of each foragers at a particular time step. The probability of death of a forager is defined to be $1-e^{-Dk}$, where $k$ is the age of the individual. A forager's vulnerability to mortality rises as he or she grows older. As a result, a forager's chances of dying grow fast. As a result, a model option was chosen such that foragers die out with a probability of $1-e^{-Dk}$. If $\mathcal{Y}$ denotes the random variable that a forager dies, then $\mathcal{Y}\sim$ Bernoulli$(1-e^{-Dk})$. So, we get two expressions of probability mass functions associated with birth and death processes as follows,
\begin{equation}
\begin{aligned}
\mathbb{P}(\mathcal{X}_n=x)&=\frac{e^\mu\mu^x}{x!}, \text{ where }\mu=mB,\text{ } x\in \mathbb{N}\\
\mathbb{P}(\mathcal{Y}=y)&=(1-e^{-Dk})^y e^{-Dk(1-y)}, \text{ where } y\in \{0,1\}.
\end{aligned}    
\end{equation}
While these processes are going on, the foragers moving in search of food may sometimes come close to the target points. How close the forager is to the target determines whether the target is achieved or not. We consider that a target point is reached if one of the foragers is at a distance lesser than $\epsilon =0.01$ from the target. 

\par Now, we construct a moving network because each time step foragers change its position (here, a forager is considered as a node). Of course, they are connected in different ways, like through agreement or communicating information to shift the colony. So, this \emph{Communication network} is implemented among the foragers in our model in the following ways.
When the foragers who moved out of the colony in search of food finds a target point, the information about the location of the achieved target point is passed on to the colony. Now for each simulation, we fix the agreement level ($p$). As in Fig.\ \ref{f1}, we can see that after getting the information, if at least $p$ fraction $(0\leq p \leq 1)$ of the static-foragers in the colony agree instantly, the colony will move, and the likelihood of this happening is $(1-p)$. As a result, there are at most $(1-p)$ fraction of the members who did not agree to relocate immediately. Now because the population's goal is to stay together and attain an ideal position from where food supplies are conveniently available, which would eventually benefit their subsequent generations, the static-foragers who had consented to migrate instantly begin convincing the rest of the foragers. This is when we introduce a penalty factor using a new parameter $\lambda$, as the static-foragers who already agreed to discuss with others in the colony in order to convince them to move. This will take some time, say $\tau$ and we assume that this convincing time to be $\big(\frac{e}{e^p}\big)^\lambda$. So we can write
\begin{equation}
\tau=\tau_0 e^{(1-p)\lambda}   
\end{equation}
where, we set $\tau_0 = 1$ and $-\infty<\lambda<\infty$. The factor $(1-p) \geq 0$ makes $0 < \tau < \infty$. This means minimum time $(\tau \approx 0)$ is required to convince when $\lambda \to -\infty$ and maximum time is required for $\lambda \to \infty$. Like birth rate and death rate, agreement level ($p$) and $\lambda$ are also fixed for a particular simulation.
 \begin{figure}
 	\includegraphics[width=0.4\textwidth]{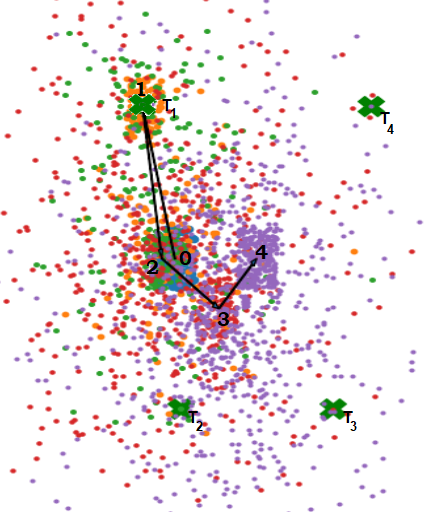}
 	\caption{{\bf Initial and final position of the colony.} There are $4$ targets ($T_1, T_2, T_3,$ and $T_4$) marked as Green crosses. The starting point of the centroid of the colony is marked as $0$ and the ending point of the centroid of it is $4$. Position $1$ is the first position of the centroid where the colony moves. Position $4$ is the best place for the colony to stay until a new target appears. The green dots represent the initial position. The orange dots
 		represent the position when the foragers find $T_1$. The red dots
 		represent the position at which the foragers discovered $T_2$. Pink
 		dots denote the position of foragers when they see target $T_3$. And
 		lastly when they discover the last target $T_4$, the position of the
 		foragers is denoted by purple.}
 	\label{f3}
 \end{figure}
 
\par When the discussion ends and the foragers admit to leaving the current location, the colony shifts to the centroid of the already searched targets.
This is the point from where the sum of the square of the distances from the already found targets is the least. We draw a diagram in Fig.\ \ref{f3} where we demonstrate the movement of colony in two dimensional space. Here the target positions are at  $(x_i,y_i)$ (where, $i=1, 2, 3, 4$) and denoted by $T_1$, $T_2$, $T_3$, and $T_4$. The targets are marked with green crosses. The entire population starts from a region whose centroid is rest at the position $0$. When, a forager has reached the first target ($T_1$), then the centroid shifts to the position$1$. The centroid follows the path: $1 \rightarrow 2  \rightarrow 3\rightarrow 4$ when any one of the foragers reaches the targets $T_1$, $T_2$, $T_3$, and $T_4$, respectively. Our objective is to find the \emph{Optimal position} for an \emph{Optimal time}, in presence of agreement level and penalty time. The detailed numerical results are described in the next section \ref{sec:results}. Note that, the trajectory of the colony movement from initial to final is not unique and it varies from one numerical experiment to another. We have numerically shown that the orientation of the colony movement does not alter the final optimal position, and optimal time does not have strong fluctuations. In this way, the colony ultimately achieves an \emph{Optimal position} at the end of the search process, and our numerical experiment terminates. 

\section{Results}
\label{sec:results}

\begin{figure}
	\includegraphics[width=0.47\textwidth]{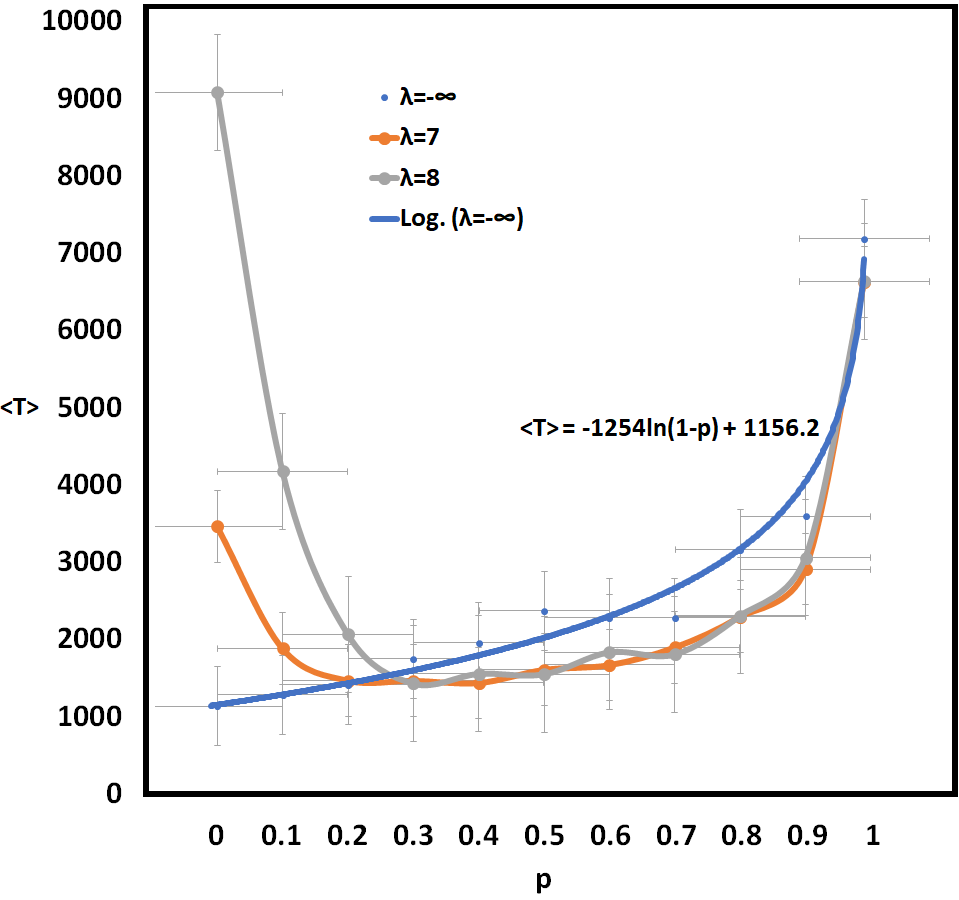}
	\caption{{\bf Variation of mean time against agreement level.} The mean time is plotted along ordinate, while the degree of agreement is depicted along abscissa, which spans from $0$ to $1$ while fixing $D/B$=0.00004. The three colors show the three different curves for different $\lambda$'s. A logarithmic trend line is fitted through the data for $\lambda=-\infty$ case with an $R^2$ value of $97.97\%$. The equation of the line is denoted in the figure.}
	\label{f4}
\end{figure}
While simulating, we tried to alter various factors like the ratio of the death rate to birth rate, $p$ and $\lambda$, which gave the following results.\\

If $T$ be the required time taken to reach the optimal position for a trial, we calculate mean time $\langle T \rangle$ over $10000$ times numerical experiments. First, we try to find a variation of mean time $\langle T \rangle$ required in order to achieve the optimal position with respect to the agreement level ($p$), keeping the penalty factor $\lambda$ and $\beta (= D/B)$ fixed and this variation is depicted in Fig.\ \ref{f4}. It is evident from this figure that $\lambda=-\infty $ with $p=0$ gives the least possible mean time as we neither require anyone to agree nor they require any time to discuss in order to move. It is as if at $p=0$ and $\lambda=-\infty$, the colony is under compulsion to move as soon as it receives information. The minima for different values of $\lambda$'s appearing at different places are nevertheless greater in magnitude than that obtained for $\lambda=-\infty$. 
Our physical perception is that it will take more time to convince more foragers and less time to convince fewer foragers. But, the figure clearly shows that for $\lambda$ equals 7 and 8, the mean time decreases almost exponentially w.ith respect to $p$ achieving a minima somewhere between $0$ and $1$ and then after a certain value of $p$, the mean time again increases. At $\lambda=-\infty$, whenever $p$ fraction of the foragers agree to move instantly, the rest have to agree instantly, and there will be no time for convincing them. At $\lambda--\infty$, we see a monotone increase in mean time $\langle T \rangle$ against $p$. At $\lambda=-\infty$ and $7$, the maximum mean time is when $p=1$ whereas for $\lambda=8$, the maxima is achieved at $p=0$. Fixing $\lambda=-\infty $, the mean time $\langle T \rangle$ keeps on decreasing and achieves minima at $p=0$. Moreover, slight fluctuations can be seen at $\lambda=-\infty$ (blue curve). However, it should have been less than the other curves as there is hardly any time spent convincing the foragers who initially didn't agree to move when the colony received the information.\\

\begin{figure}
	\includegraphics[width=0.47\textwidth]{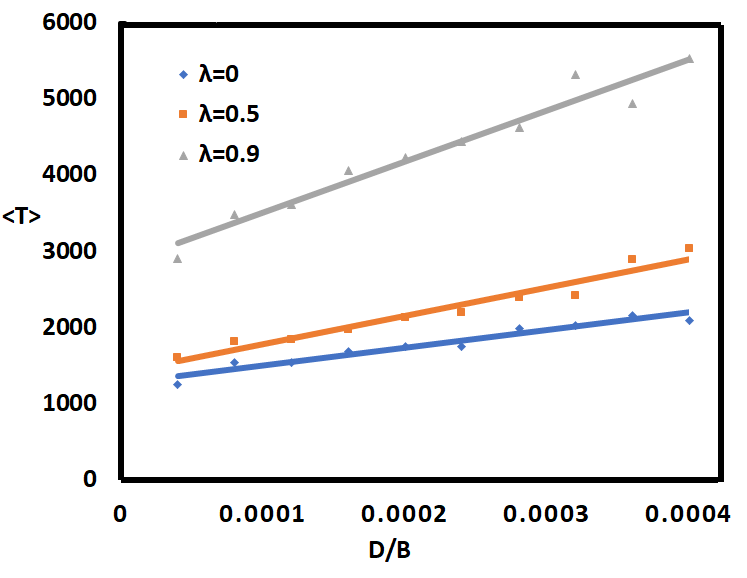}
	\caption{{\bf Mean time versus Death-to-Birth ratio.}
	The blue points represents $\lambda$ is 0, whereas the orange graph represents $\lambda$ is 0.5 and the grey graph represents $\lambda$ is 0.9. The agreement level is fixed at 0.6, and three linear lines have also been fitted to the data with slopes $7\times10^6, 4\times10^6, 2\times10^6$ respectively.}
	\label{f5}
\end{figure}

\par
Figure\ \ref{f5} depicts the variation of mean time ($\langle T \rangle$) with respect to the ratio of Death Rate to Birth Rate, setting Penalty factor $\lambda$ to be $0.0$, $0.5$, and $0.9$ fixing the agreement level ($p$). The ratio on the $x$-axis ranges from $0$ to $0.0004$, with a typical time of $1000$ to $6000$ in the Y-axis. At each values of $\lambda$, the graph shows a nearly identical fluctuation in $\langle T \rangle$ in relation to the ratio. The graph shows an increasing linear trend in $\langle T \rangle$ as the ratio of death to birth is increased. The graph also shows that the $\langle T \rangle$ for $\lambda=0.9$ is greater than that for $\lambda=0.5$ at all values of Death to Birth ratio. Similarly, $\langle T \rangle$ for $\lambda=0.5$ is greater than that for $\lambda=0$ at all values. Later examination showed that the regression lines have $R^2$ accuracy of $94.3\%,94.92\%,95.12\%$ respectively. It is also clear from the graph that the slopes of the fitted line at $\lambda=0.9$ is greater than that of the fitted line at $\lambda=0.5$, which is again greater than the slope of the fitted line at $\lambda=0$. Therefore, the rate of increase of $\langle T \rangle$ against $(D/B)$ at $\lambda=0.9$ is greater than that of the rate of increase of $\langle T \rangle$ against $(D/B)$ at $\lambda=0.5$, which is again greater than the rate of increase of $\langle T \rangle$ against $(D/B)$  at $\lambda=0$.

\begin{figure}
	\includegraphics[width=0.45\textwidth]{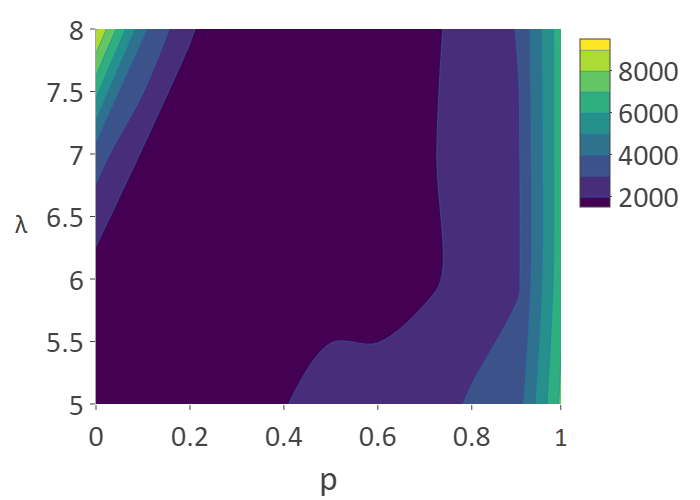}
	\caption{{\bf Mean time with variation of $p$ and $\lambda$}. A contour plot of mean time as a function of $p$ and $\lambda$, i.e., $\langle T \rangle = f(p,\lambda)$.}
	\label{f6}
\end{figure} 

\par Figure\ \ref{f6} is a contour plot which depicts the variation of $\langle T \rangle$ with respect to the variation of $p$ and $\lambda$ simultaneously. The contour plot also displays an intriguing model concept. There is a kink in the line $\lambda=6.5$, as seen. This demonstrates that the time spent in making a decision is so minimal that it barely hinders the process of getting to the best position. 
Moreover, The contour shows that for values of $p$ between $0$ and $0.8$ and for most of the values of $\lambda$, mean time required is less than $2000$. There is no value of $\lambda$ for which mean time is lesser than $4000$ for $p\in(0.9, 1)$. Thus when agreement is $100\%$, for all values of the penalty factor, the mean time taken always remains higher than $4000$.

\section{Discussion}
\label{sec:disc}
As we already know that foraging is the process of looking for food resources. It has an impact on an animal's fitness since it is crucial to an animal's capacity to live and reproduce. To understand foraging, behavioral ecologists utilize economic models and categories; many of these models are optimal models. Thus, foraging theory is explained in maximizing a foraging decision's reward. Many of these models' payout is the amount of energy the animal obtains per unit of time, especially the highest ratio of energetic gain to cost during foraging.

\par Foraging can be categorized into two main types. The first is solitary foraging when animals forage by themselves. The second is group foraging. Group foraging includes when animals can be seen foraging together, when it is beneficial for them to do so, and when it is detrimental for them to do so. The fundamental goal of this work is to build a model in which we attempt to include group foraging, which is advantageous most of the time, particularly when there is no rivalry among the species. And in our model, we expect it to be this because there is no competition, and therefore we strive to build an optimum technique for obtaining the optimal position in the shortest amount of time.
\par It is always our physical experience that when the amount of agreement is high, the required time is reduced.  However, the penalty factor is crucial in a population model like the one we attempted to create. For each penalty factor, we find some ideal agreement level where the time required is smaller and which is not always the $100\%$ agreement level. For some values of penalty factor, we even get non-monotonous behavior in the mean time with a rise in agreement level, with minima in between and subsequently an increase. When the amount of agreement is low, i.e., $p$ is near $0$, the colony is more likely to spend more time at the initial site depending on the value $\lambda$ takes. This inherent delay accumulates, resulting in a substantially longer period to attain an ideal posture. When the $p$ is near $1$, however, it takes significantly longer to discuss, delaying the process of finding an ideal site. Therefore, there are some optimal strategies to obtain the optimal position depending on the value of the penalty factor ($\lambda$).
\par Then, our emphasis changes to determine the best approach depending on the Death-to-Birth ratio ($\beta$). When the penalty factor is set to $0, 0.5,$ and $0.9$, the mean time varies almost linearly for Death-to-Birth ratio $(\beta)$'s ranging from $0$ to $0.004$. For all $\lambda$ values, the mean time increases approximately linearly and reaches a maximum when the ratio is equal to $0.0004$. We also see that the mean time taken to obtain the optimal position at $\lambda=0.9$ is greater than mean time at $\lambda=0.1$ is greater than mean time at $\lambda=0$. This suggests that increasing the penalty factor value can increase the mean time required. When other criteria such as agreement level ($p$), Death-to-Birth ratio ($D/B$), and penalty factor ($\lambda$) are held constant, a lower penalty factor is preferred. 
We also come to know that for the values of penalty factor taking between $0$ and $8$, the mean time for agreement level close to $1$ remains greater than $6000$. But there are also agreement levels that take even lesser than $3000$ to reach the optimal position. This is a clear indication of the fact that there is a better strategy to achieve the optimal position in minimum time. Thus, altering one parameter while holding the others constant always yields an optimum approach. Using these strategies, we can reduce the time of achieving the optimal position, which will be ultimately beneficial for the population. 
\par The future aspect of this type of research is reducing the time spent searching for targets when drones are utilized. One of the most critical responsibilities in drone operations is finding a target rapidly. Rapid target detection is very important for jobs like detecting rescue victims during the golden period, monitoring the environment, detecting military sites, and monitoring natural catastrophes.

\section*{Authorship contribution statement}
{\bf Sanchayan Bhowal \& Ramkrishna Jyoti Samanta}: Conceptualization; Methodology; Software; Validation; Investigation; Writing - original draft;
Visualization; {\bf Arnob Ray Sirshendu Bhattacharyya \& Chittaranjan Hens}: Supervision, Conceptualization, Writing – review \& editing.

\section*{Data availability}
All codes used in this study are made publicly available at \url{https://github.com/Sanchayan-Bhowal/Forager}.

\section*{Appendix}
\label{app}
Hence, the population becomes random depending on the birth and death rates fixed for a particular simulation.
Let us define some random variables which will help us measure the population growth of the foragers.

\begin{equation}
\begin{aligned}
Z_n & \coloneqq \text{No. of foragers at }n^{\text{th}} \text{ step}\\
\chi_{n,k} & \coloneqq \text{No. of foragers at }n^{\text{th}} \text{ step of age } k\\
\chi_{n} & \coloneqq \{\chi_{n,k}\}_{k=1}^{\infty} \text{i.e., a sequence of }\chi_{n,k}\\
Y_{n,k} & \coloneqq \text{No. of foragers of age k who dies at }n^{\text{th}} \text{ step}\\
& ~~~~~~~~~~~~~~~~~~~~~~~~~~~~~~~~~~~~~~~~~~~~~~~~~~~~~~~~~~~~~~~~\hfill \forall \; k\in \mathbf{N}
\end{aligned}
\end{equation}

The distributions of these random variables depend on the previous step. However, the conditional distribution can be determined. The distribution of $\chi_{n+1,1}|\chi_n$ is given by Poisson distribution with parameter $\mu= m$\emph{B}, where $B$ is the Birth rate, and $m$ is the number of foragers who are capable of reproducing. Hence, $m=\sum_{k=20}^{60}\chi_{n,k}$ foragers are aged between $20$ and $60$, therefore can reproduce. On the other hand, each of the foragers of age $k$ can die with probability $1-e^{-Dk}$. Hence, the number of foragers dying in a particular age group is given my Binomial with $\chi_{n,k}$ as the number of foragers (by definition).
This is described as below

\begin{equation}
\begin{aligned}
\chi_{n+1,1}|\chi_n & \sim \text{Poisson}(B(\sum_{k=20}^{60}\chi_{n,k}))\\
Y_{n,k}|\chi_n & \sim   \text{Binomial}(\chi_{n,k},1-e^{-Dk})\\
\end{aligned}
\end{equation}

The definitions of the random variables give rise to some useful relations. The number of foragers of age $k$ in $(n+1)^{\mathrm{th}}$ step is given by the foragers who were of age $k-1$ in $n^{\mathrm{th}}$ and then removing those who died in the previous step. This relation also expresses the conditional expectation with respect to $\chi_{n,k-1}$.

\begin{equation}
\begin{aligned}
\chi_{n+1,k}&=\chi_{n,k-1}-Y_{n,k-1}\\
E[\chi_{n+1,k}|\chi_n]&=E[\chi_{n,k-1}-Y_{n,k-1}|\chi_n]\\
&=E[\chi_{n,k-1}|\chi_n]-E[Y_{n,k-1}|\chi_n]\\
&=\chi_{n,k-1}-\chi_{n,k-1}(1-e^{-D(k-1)})\\
&=\chi_{n,k-1}e^{-D(k-1)}
\end{aligned}
\end{equation}
The total number of foragers in the $(n+1)^{\mathrm{th}}$ step,$Z_{n+1}$ is given by sum of $\chi_{n,k}$ over all age groups. Conditioning on $\chi_n$ we can get the conditional expectation of $Z_{n+1}$ as shown in Eq.\ \ref{eq:Z}.
\begin{equation}
\label{eq:Z}
\begin{aligned}
Z_{n+1} &=\sum_{k=1}^{\infty}\chi_{n+1,k}\\
E[Z_{n+1}|\chi_n] &=\sum_{k=1}^{\infty}E[\chi_{n+1,k}|\chi_n]\\
&=E[\chi_{n+1,1}|\chi_n]+\sum_{k=2}^{\infty}\chi_{n,k-1}e^{-D(k-1)}\\
&=B(\sum_{k=20}^{60}\chi_{n,k})+\sum_{k=2}^{\infty}\chi_{n,k-1}e^{-D(k-1)}
\end{aligned}
\end{equation}

The Eq.\ \ref{eq:Z} shows that once the $\chi_{n,k}=0~\forall k\leq60$, then no more foragers are born neither any of the foragers with the ability to reproduce are left. In that case the old foragers perish as time grows, which is evident from the factor of $e^{-D(k-1)}$ which tends to 0 as $k$ increases. This phenomena where the species die out on long run is defined to be \emph{Population Collapse}. Such a phenomena is highly criticized as it is not beneficial for the species to survive.




\end{document}